\documentclass[aps,prl,showpacs,twocolumn,superscriptaddress,bibnotes]{revtex4}

\usepackage{graphicx}
\begin{document}

\newcommand{\Hpip}{(C$_5$H$_{12}$N)$_2$CuBr$_4$}
\newcommand{\Hpips}{(Hpip)$_2$CuBr$_4$}
\newcommand{\Dpip}{(C$_5$D$_{12}$N)$_2$CuBr$_4$}
\newcommand{\Dpips}{(Dpip)$_2$CuBr$_4$}

\title{Field--Controlled Magnetic Order in the Quantum Spin--Ladder System 
(Hpip)$_2$CuBr$_4$}

\author{B.~Thielemann}
\affiliation{Laboratory for Neutron Scattering, ETH Zurich and Paul Scherrer
Institute, CH--5232 Villigen, Switzerland}

\author{Ch.~R\"uegg}
\affiliation{London Centre for Nanotechnology, University College London,
London WC1E 6BT, United Kingdom}

\author{K.~Kiefer}
\affiliation{BENSC, Helmholtz Centre Berlin for Materials and Energy, D--14109
Berlin, Germany}

\author{H. M.~R{\o}nnow}
\affiliation{Laboratory for Quantum Magnetism, Ecole Polytechnique
F\'ed\'erale de Lausanne, CH--1015 Lausanne, Switzerland}

\author{B. Normand }
\affiliation{Theoretische Physik, ETH--H\"onggerberg, CH--8093 Z\"urich,
Switzerland }

\author{P.~Bouillot}
\affiliation{DPMC-MaNEP, University of Geneva, CH--1211 Geneva, Switzerland}

\author{C.~Kollath}
\affiliation{Centre de Physique Th\'eorique, Ecole Polytechnique, CNRS, 91128
Palaiseau Cedex, France}

\author{E.~Orignac}
\affiliation{LPENSL CNRS UMR 5672, F-69364 Lyon Cedex 07, France}

\author{R.~Citro}
\affiliation{Dipartimento di Fisica "E. R. Caianiello" and CNISM, Universit\`a
di Salerno, I--84100 Salerno, Italy}

\author{T.~Giamarchi}
\affiliation{DPMC-MaNEP, University of Geneva, CH--1211 Geneva, Switzerland}

\author{A. M.~L\"auchli}
\affiliation{Institut Romand de Recherche Num\'{e}rique en Physique des
Mat\'{e}riaux (IRRMA), CH--1015 Lausanne, Switzerland}

\author{D.~Biner}
\author{K. Kr\"amer}
\affiliation{Department for Chemistry and Biochemistry, University of Bern,
CH--3000 Bern 9, Switzerland}

\author{F.~Wolff--Fabris}
\author{V.~Zapf}
\author{M.~Jaime}
\affiliation{MPA-NHMFL, Los Alamos National Laboratory, Los Alamos, New Mexico
87545, USA}

\author{J.~Stahn}
\affiliation{Laboratory for Neutron Scattering, ETH Zurich and Paul Scherrer
Institute, CH--5232 Villigen, Switzerland}

\author{N. B.~Christensen}
\affiliation{Laboratory for Neutron Scattering, ETH Zurich and Paul Scherrer
Institute, CH--5232 Villigen, Switzerland}
\affiliation{Ris{\o} National Laboratory for Sust. Energy,~Technical
University of Denmark, DK-4000 Roskilde}

\author{B. Grenier}
\affiliation{Universit\'e Joseph Fourier, Grenoble and CEA-Grenoble,
INAC/SPSMS/MDN, F--38054 Grenoble, France}

\author{D. F.~McMorrow}
\affiliation{London Centre for Nanotechnology, University College London,
London WC1E 6BT, United Kingdom}

\author{J.~Mesot}
\affiliation{Laboratory for Neutron Scattering, ETH Zurich and Paul Scherrer
Institute, CH--5232 Villigen, Switzerland}
\affiliation{Laboratory for Quantum Magnetism, Ecole Polytechnique
F\'ed\'erale de Lausanne, CH--1015 Lausanne, Switzerland}

\date{\today}

\begin{abstract}

Neutron diffraction is used to investigate the field--induced, 
antiferromagnetically ordered state in the two--leg spin--ladder material 
\Hpips. This "classical" phase, a consequence of weak interladder coupling,
is nevertheless highly unconventional: its properties are influenced strongly 
by the spin Luttinger--liquid state of the ladder subunits. We determine 
directly the order parameter (transverse magnetization), the ordering 
temperature, the spin structure, and the critical exponents around the 
transition. We introduce a minimal, microscopic model for the interladder 
coupling and calculate the quantum fluctuation corrections to the 
mean--field interaction.

\end{abstract}


\maketitle
                       
Low--dimensional magnets have been the subject of intense theoretical
research for many decades. Of particular interest are the intriguing
ground-- and excited--state properties of one--dimensional (1D) systems
such as chains and ladders \cite{Bethe31,Haldane83,Dagotto95,Sachdev94,
Chitra,furusaki,Giamarchi99}. In this context, residual interactions
between the low--dimensional units, which are always present in real
materials, may be viewed as a distraction from the intrinsic physics.
However, such interactions open up fascinating new avenues of 
investigation concerning the crossover from one-- to higher--dimensional
behavior. In 3D, antiferromagnetic (AF) magnons in a gapped, quantum
magnet undergo Bose--Einstein Condensation (BEC) at a magnetic field
$B_c$, where the gap is closed by the Zeeman effect \cite{Giamarchi08}.
At this quantum critical point, the spin components develop long--ranged
order perpendicular to the magnetic field (3D--XY type). By contrast, in
1D any long--range order is destroyed by quantum phase fluctuations and a
critical phase with algebraic spin correlations -- a spin Luttinger liquid
(LL) -- is predicted \cite{Sachdev94,Giamarchi99}. While the spin LL may 
be realized at finite temperatures in coupled $S=1/2$ chain systems such
as KCuF$_3$ \cite{Schulz96,Lake05}, a particularly rich phase diagram is
expected for weakly coupled ladders: here the spin LL is induced from a
gapped, quantum disordered (QD) phase by an applied magnetic field, and
its LL parameters can be tuned directly by the field \cite{Giamarchi99}.

\begin{figure}[t!]
\begin{center}
\includegraphics[width=0.45\textwidth]{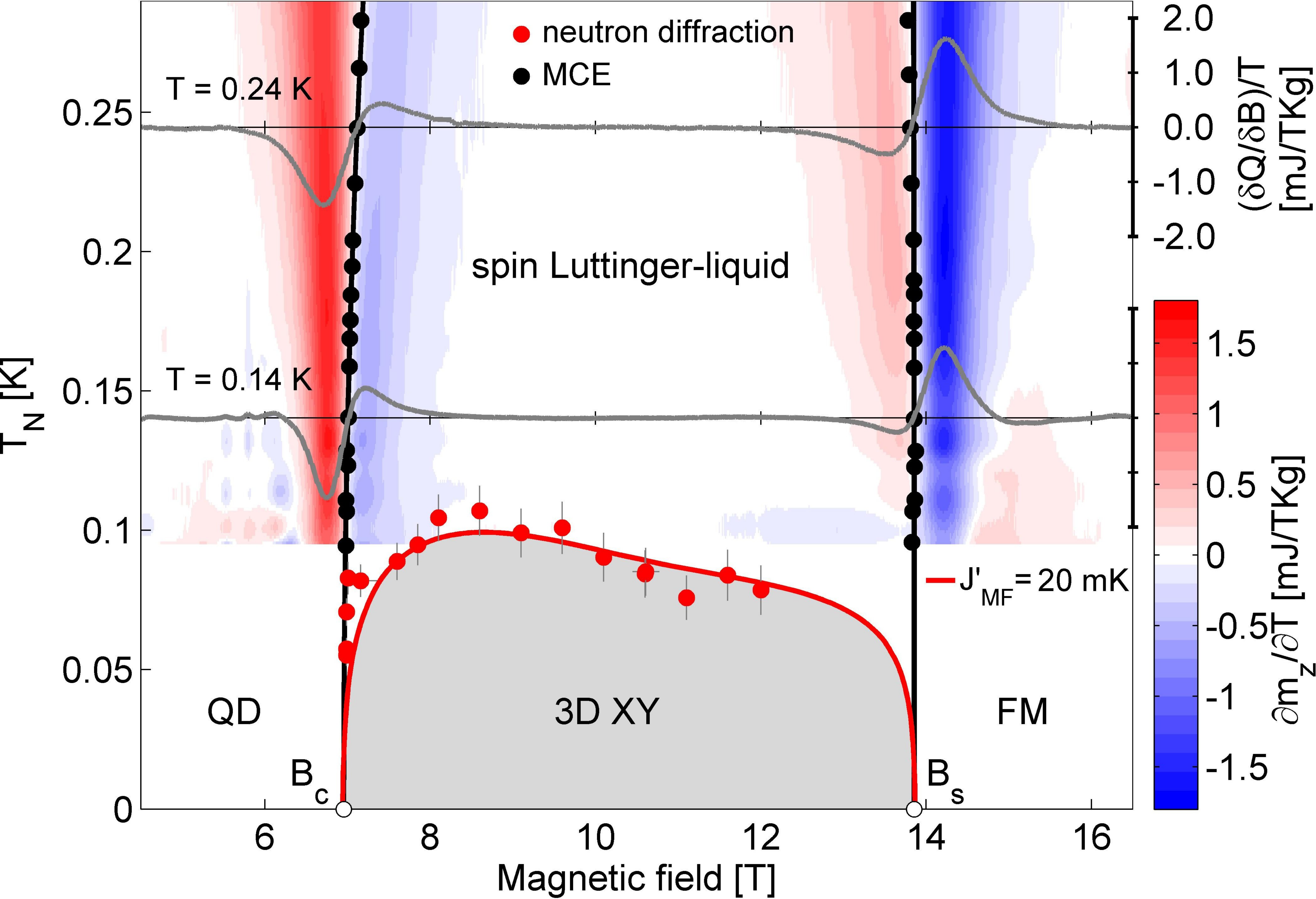}
\end{center}
\caption{\small Low--$T$ phase diagram of (Hpip)$_2$CuBr$_4$. 
The crossover temperature to the spin LL phase is derived from
MCE measurements and the phase transition to the BEC (3D--XY magnetic order)
from neutron diffraction. The contour plot is based on 18 individual field
scans of the MCE (two shown as gray lines), using $(\delta Q/\delta B)/T
 = -(\partial m_z/\partial T)|_{B}$ \cite{Ruegg_specificheat}. The red line 
is based on a theoretical fit (see text).}
\label{fig1}
\end{figure}

Materials realizing quasi--1D spin--ladder geometries, and with critical
fields $B_c$ (QD to LL) and $B_s$ (magnetic saturation) accessible in the
laboratory, are rare \cite{Dagotto95}. While the latter difficulty is 
generally overcome in metal--organic compounds, these can still suffer 
from other complications, such as the additional terms found in the 
magnetic Hamiltonian of CuHpCl \cite{B_CuHpCl}. Thus the system \Hpip~
(\Hpips) \cite{Patyal,Watson} is unique in its class: numerous thermodynamic
measurements are in quantitative agreement with predictions for an ideal
ladder \cite{Lorenz,Ruegg_specificheat,Berthier_NMR}, and inelastic
neutron scattering (INS) demonstrates a high degree of one--dimensionality
\cite{Thielemann}. All of these techniques point to a minimal Heisenberg
ladder Hamiltonian with respective rung and leg exchange constants
$J_r = 12.9(2)$ K and $J_l = 3.3(3)$ K. Only nuclear magnetic resonance
(NMR) measurements have to date been performed at temperatures low 
enough to access the energy scale of the interladder coupling: in
Ref.~\cite{Berthier_NMR}, the transition to 3D order is found at and below
110 mK, and the phase boundary interpreted in terms of a field--tuned spin
LL regime between $B_c$ and $B_s$.

In this Letter we report the results of comprehensive neutron diffraction
studies, performed at dilution--refrigerator temperatures, of the 3D ordered 
phase in 
\Hpips. We determine the spin structure and measure directly the transverse 
magnetic moment as a function of field and temperature. The vanishing of 
the order parameter, combined with the magnetocaloric effect (MCE), give 
the LL crossover, the LL exponents close to the 3D regime, and independent 
measurements of the phase boundary. From the spin structure we deduce 
a microscopic model for the interladder coupling which allows a quantitative 
determination of the interaction between the 1D subunits, using a combination 
of mean--field calculations \cite{Berthier_NMR} and Quantum Monte Carlo (QMC) 
simulations of the full interacting--ladder model.

The MCE was measured on high--quality single--crystalline \Hpips~in a 
standard dilution refrigerator at the NHMFL in Los Alamos, with sweep 
rates between 0.025 T/min and 0.075 T/min. Neutron diffraction experiments 
were performed on deuterated \Dpips~single crystals with sample mass 200 mg 
on the instruments D23 at the ILL and RITA--2 at SINQ (PSI), using standard 
set--ups. For all measurements, a vertical magnetic field was applied along 
the crystallographic $b$--axis, {\it i.e.}~perpendicular to the ladders.

Before studying the 3D ordered phase, it is necessary to understand in
full detail the disordered phase from which it emerges. The quasi--1D regime 
above the 3D phase boundary is investigated using the MCE, which maps the 
crossover from the QD state into the spin LL through local extrema in the 
temperature--dependence of the longitudinal magnetization $m_z$. Figure 1 
shows $\partial m_z/\partial T$ (contour plot) and these extrema (black 
circles) down to 100 mK, extending (from 300 mK) the results of our previous 
measurements \cite{Ruegg_specificheat} and providing a frame of reference for 
other reported results \cite{Berthier_NMR}. The LL crossover is analyzed 
by a sliding--window technique \cite{Nohadani}, whose first step is the 
determination of the critical fields $B_{c} = 6.96(2)$ T and $B_{s} =  
13.85(3)$ T. With these values fixed, the crossover temperature is fitted 
to $T_{\rm LL} \propto (B-B_{c})^{\frac{1}{\nu}}$ (black lines), yielding 
an exponent $\nu = 2.1(1)$ at $B_{c}$ \cite{nu_hs}. A ladder spin system 
in a field $B \approx B_c$ can be mapped to a free--fermion model (LL 
exponent $K = 1$), and hence one expects $\nu = 1$ \cite{Chitra}. However, 
a bosonization interpretation
of density--matrix renormalization--group (DMRG) calculations for the
ladder model \cite{Berthier_NMR} shows that $K(B)$ decreases rapidly
below 1 as the field is moved away from $B_c$ and $B_s$. Consequently,
the true critical regime is very narrow, a result seen also in QMC
calculations for the related Haldane spin chains \cite{Oshikawa}. Thus
while our MCE measurements demonstrate spin LL behavior down to 100 mK,
the universal exponents are not reached before the LL regime is cut off
by 3D order. Close to the critical fields, our results indicate that the
ladder system remains far from this universal regime even for $T/J_l
\approx 0.03$.

\begin{figure}[t!]
\begin{center}
\includegraphics[width=0.45\textwidth]{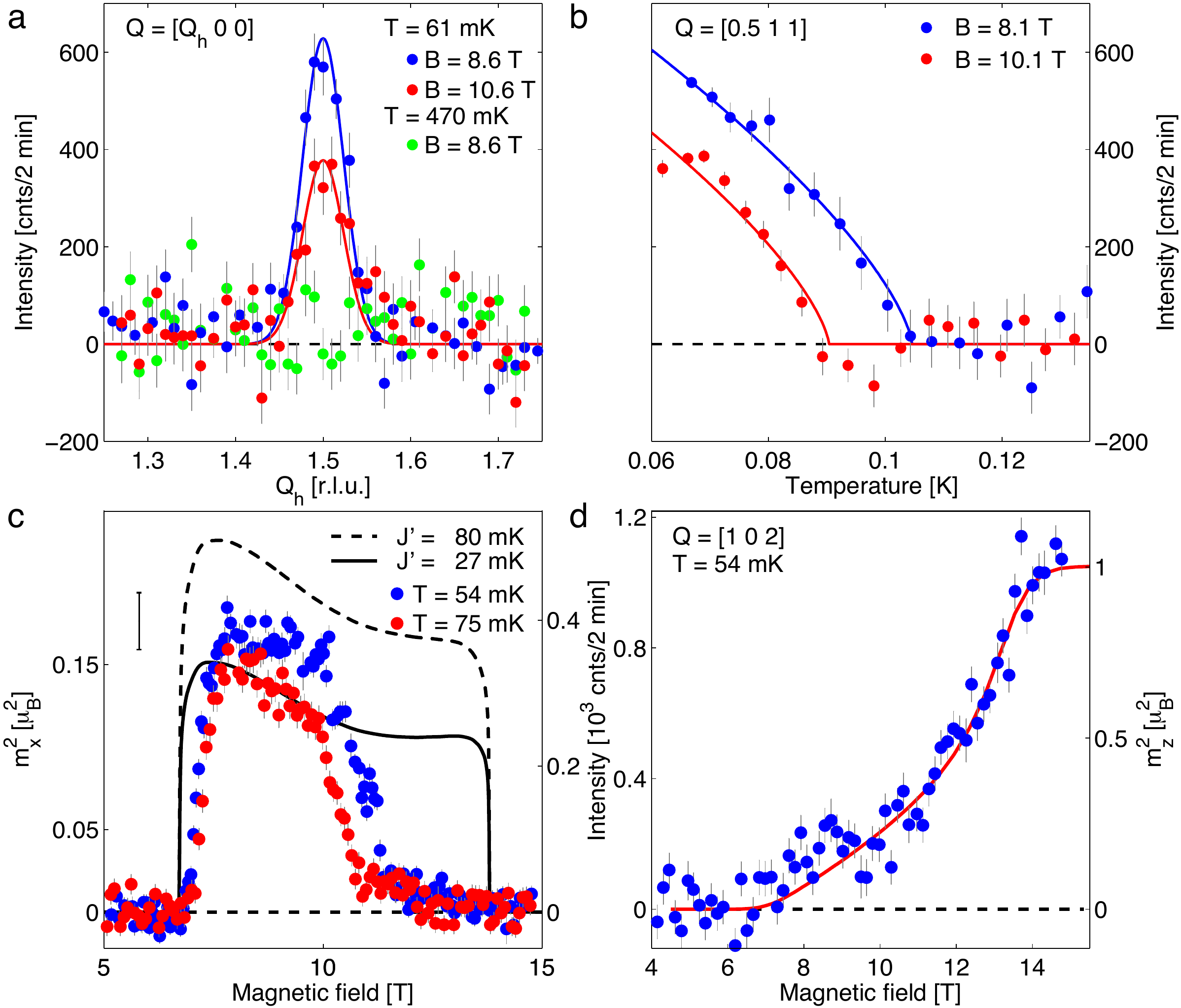}
\end{center}
\caption{\small Summary of neutron diffraction data. (a) $Q$--scans across
an AF Bragg peak after subtraction of a flat background measured in the QD
phase at $B = 6$ T and $T = 63$ mK. (b) $T$--dependence of the Bragg
intensity, demonstrating the onset of 3D long--range order at $T_{\rm N}(B)$;
solid lines are fits using the 3D--XY exponent. (c) $B$--dependence of
$m_x^2$, measured at $Q=$(1.5~0~0) for $T =$ 54 mK (blue) and $T =$ 75 mK
(red). Solid and dashed lines are from DMRG MFA for the given interladder
interaction $J'$. Error bars on data points are based on counting statistics,
while the vertical black line indicates the systematic error of the
calibration to absolute units. (d) Magnetic signal at $Q=$(1~0~2), which
is proportional to $m_z^2$: shown is the neutron intensity after subtraction
of the nuclear contribution, which also corrects for magnetostriction effects.
The red line is obtained from a QMC calculation.}
\label{fig2}
\end{figure}

Turning now to the ordered phase, Fig.~2 summarizes the results of our
neutron diffraction measurements, which were taken at temperatures down
to 54 mK. Figure 2(a) shows $Q$--scans across the AF wave vector,
$Q=$(1.5~0~0), on cooling the sample from the spin LL regime at $B_c <
B < B_s$. Resolution--limited magnetic Bragg peaks are observed at
base temperature, demonstrating long--range AF order. The magnetic
Bragg peak remains at the same commensurate position, but its
intensity decreases with increasing field, indicating a substantial
field--dependence of the transverse ordered moment. The
temperature--dependence of the Bragg intensity is presented in
Fig.~2(b): the vanishing of the magnetic signal is used to determine
the phase boundary shown as red circles in Fig.~1, and its thermal
evolution is very well described by the critical exponent, $2\beta =
0.70$, of the 3D XY model \cite{Giamarchi08} over the full range of data
available (essentially $\frac12 T_{\rm N} < T < T_{\rm N}$).

Figure 2(c) shows field scans of the AF Bragg intensity at $Q=$(1.5~0~0)
for $T = 54$ mK and $T = 75$ mK. This is proportional to the square of
the transverse magnetization, $m_x^2$, and was scaled to the ordered
moment obtained at $B = 8.6$ T from a complete refinement of the spin
structure (below). In contrast to NMR, $m_x^2$ is determined directly
by neutron scattering, allowing additional quantitative tests of
theoretical predictions. From Fig.~2(b), the AF order parameter is not 
saturated at $T = 54$ mK, and no Bragg peak (or 3D transition) is observed 
for $B > 12$ T. Above $B_c$, the uniform magnetization $m_z$ increases 
monotonically, which generates a small, ferromagnetic (FM) signal on top 
of the nuclear Bragg peaks. Our characterization of all components of the 
magnetization is completed by field scans of this magnetic intensity, as 
shown in Fig.~2(d) for ${\mbox{\boldmath$Q$\unboldmath}} =$(1~0~2). The 
red line is obtained from a QMC calculation of the ladder magnetization, 
$m_z^2$, using the exchange interactions cited above, and shows good 
overall agreement.

\begin{figure}[t!]
\begin{center}
\includegraphics[width=0.45\textwidth]{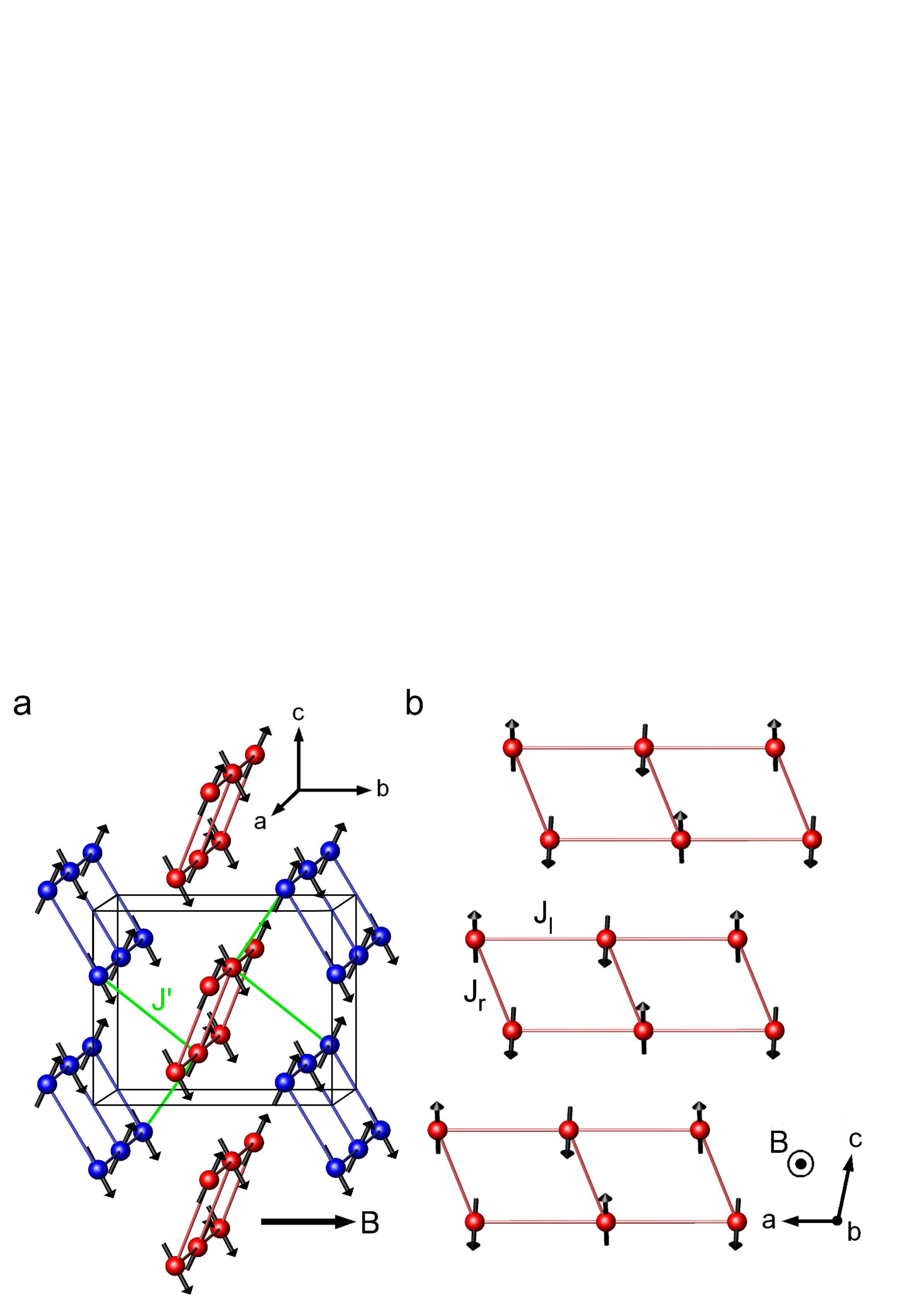}
\end{center}
\caption{\small Magnetic structure (black arrows) in the 3D ordered phase
of \Dpips, determined by neutron diffraction at $B=8.6$ T and $T=63$ mK.
Only Cu atoms forming the ladders are shown (red and blue). The spin
component along the field ($B||b$) is fixed by QMC calculations.
Projections are shown (a) on the $bc$--plane and (b) on the $ac$--plane.
Interladder bonds $J'$ for one rung are shown in green.}
\label{fig4}
\end{figure}

In addition to $m_z$ and $m_x$, neutron diffraction also allows a 
quantitative determination of the magnetic structure. At base temperature 
and $B = 8.6$ T (maximum $T_{\rm N} \approx 110$ mK), the intensities of 
26 AF Bragg peaks were recorded on D23. Among the four allowed magnetic
structures, that shown in Fig.~3 provides the best fit ($\chi^2 = 5.54$):
the spins are aligned perpendicular to the $a$--axis and antiparallel
within the ladder, but parallel on ladders of the same type [propagation 
vector ${\mbox{\boldmath$k$\unboldmath}} =$ (0.5 0 0)]. The ordered moment 
is 0.41(2) $\mu_B$ per copper ion. Its orientation is parallel to the 
maximum of the $g$--factor in the $ac$--plane \cite{Patyal}. We note that 
nearest--neighbor Cu atoms between adjacent ladders of opposite type 
(red and blue) have FM aligned spins [Fig.~3(a)].

A qualitative discussion of our results is aided by a specific model for
the interladder coupling responsible for 3D order. Here we restrict our
considerations to a minimal model with only one interladder interaction
parameter, of magnitude $J' = {\rm O}(100$ mK). Inspection of the lattice
structure of \Hpips, which is similar to that of TlCuCl$_3$ (albeit with
the dimer units rather less tilted relative to the ladder axis), motivates 
the choice of the bond $J_{3}$ in the notation of Ref.~\cite{Matsumoto04}.
This bond connects ladders of opposite orientations [red to blue in 
Fig.~3(a)] along the directions $\pm$(1 $\pm$0.5 0.5), meaning that the 
Cu sites are displaced by one unit along the $a$--axis. AF bonds $J' 
\equiv J_3$ then ensure the FM alignment mentioned above, and a completely
unfrustrated spin structure. Any one ladder rung has four bonds of this
type (coordination $z = 4$). INS measurements of the very small triplet
dispersions along the $b$ and $c$ axes tend to support this type of model;
full details of the 3D spin dynamics and exchange paths will be presented
elsewhere \cite{Thielemann}.

The measured transition temperature $T_{\rm N}(B)$ is shown in Fig.~1. The 
marked asymmetry about $B = (B_c + B_s)/2$ ($m_z = 1/2$) arises from the 
changing influence of the upper two triplet branches as the field is 
increased, and is in strong contrast to spin systems in two and three 
dimensions, such as BaCuSi$_2$O$_6$ and TlCuCl$_3$ \cite{Giamarchi08}. This 
asymmetry is also apparent in QMC studies of coupled ladders \cite{Wessel}. 
The solid red line in Fig.~1 is fitted from the theory presented in 
Ref.~\cite{Berthier_NMR}: $T_{\rm N}(B)$ depends on the LL parameters of 
the ladders, which are themselves functions of $B$ and are obtained from 
a bosonization interpretation of DMRG calculations performed for a ladder 
with $J_r/J_l = 3.6$. Interladder coupling is treated in a mean--field 
approximation (MFA), and the only free parameter in Fig.~1 is the 
magnitude of this exchange constant, $J'_{\rm MF}$. The best fit to our
experimental data yields $J'_{\rm MF} = 20(1)$ mK, providing valuable
independent confirmation of the results of Ref.~\cite{Berthier_NMR}.

\begin{figure}[t!]
\begin{center}
\includegraphics[width=0.48\textwidth]{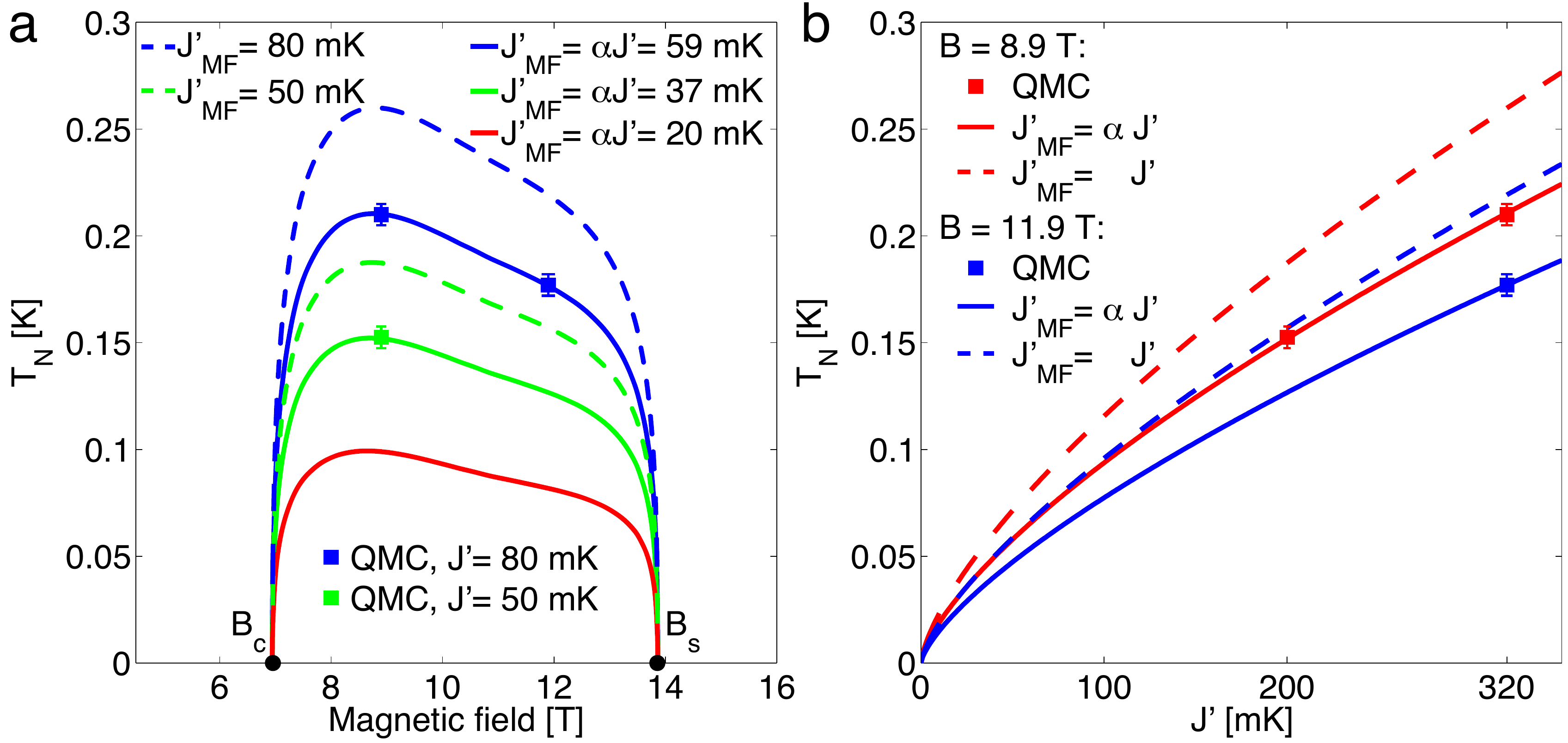}
\end{center}
\caption{\small The 3D--XY phase boundary by DMRG MFA and QMC. 
(a) $T_{N}(B)$ for several values of $J'$, as indicated, $\alpha$
 = 0.74(1). (b) $T_{N}$ as a function of $J'$.}
\label{fig3}
\end{figure}

However, by its nature the MFA underestimates the effects of quantum
fluctuations, and thus overestimates the value of $T_{\rm N}$ corresponding 
to a given $J'$ (for coupled 1D chains by approximately 50\% \cite{Schulz96}). 
We have obtained the "bare" value of $J'$ by performing QMC simulations 
\cite{ALPS,ALPS_SSE} for the interladder exchange geometry of the minimal 
model, which is a very demanding computational task due to the strong 
spatial anisotropy of the exchange constants.
It has been shown numerically \cite{Yasuda05} that quantitative
agreement of mean--field and exact results can be recovered simply by
introducing a renormalization factor $\alpha$, which is a constant in
the weak--coupling limit, and for the purposes of the current analysis
can be viewed as a renormalization of the coupling $J'_{\rm MF} = \alpha J'$.
The 3D ordering temperature computed by QMC is shown in Fig.~4(a) at $B
 = 8.9$ T (and 11.9 T) for two chosen values of $J'$; the solid 
lines are obtained from the DMRG MFA. As shown in Fig. 4(b), a single value 
$\alpha$ = 0.74(1) is indeed sufficient to ensure perfect agreement between 
the QMC simulations and the DMRG MFA. Hence the best fit to $J'_{\rm MF}$ 
corresponds to a bare (microscopic) coupling $J'$ = 27(2) mK.

A further test of the coupling model and the DMRG MFA is provided by the
transverse--moment measurements shown in Fig.~2(c). The qualitative shape
of the zero--temperature DMRG--based results is again asymmetric, due to
the changing influence of the high--lying triplet modes and to the
field--dependence of the underlying LL parameters. While this appears 
to mirror the data, a quantitative comparison faces two complications: 
experimentally, the data is not in the low--temperature limit, and indeed 
falls unexpectedly above 10.5 T; theoretically, the calculated moment is 
rather insensitive to the value of $J'$, while the quantum fluctuation 
suppression factor for $m_x^2$, which need not match that deduced from 
$T_{\rm N}$, is unknown. Clearly even the unrenormalized value, $J' = 27$ 
mK, somewhat underestimates the transverse moment, and significantly higher 
values, up to $J' \approx 80$ mK (Fig.~2(c), \cite{Thielemann}), appear to 
be required in this framework. More detailed experiments and theoretical 
analysis will be necessary to resolve this discrepancy.

In summary, comprehensive neutron diffraction data and measurements of the
magnetocaloric effect are presented to investigate the 3D ordered phase
realized in the prototypical spin--ladder material \Hpips~at low temperatures
and high magnetic fields. We determine the temperature of the transition
which separates 1D from 3D physics in coupled spin Luttinger liquids, and
characterize the critical behavior in the 1D regime. In the 3D phase,
we measure both the transverse and longitudinal magnetizations, and
establish the spin structure. The unconventional field--dependences of
the N\'{e}el temperature and of the transverse magnetization agree well 
with a description based on a minimal coupling model and combining DMRG 
calculations with a mean--field treatment of the (renormalized)
interladder interactions. We determine the renormalization factor for
this coupled--ladder geometry by comparison with detailed QMC simulations.

We thank C. Berthier for valuable discussions. This project was supported
by the Swiss National Science Foundation through the NCCR MaNEP and Division
II, by the Royal Society, EPSRC, NSF, DOE, the RTRA network "Triangle de la
Physique", and the State of Florida through the National High Magnetic Field
Laboratory. The work is based in part on experiments performed at the Swiss
spallation neutron source, SINQ, at the Paul Scherrer Institute, Villigen,
Switzerland.


\begin{thebibliography}{00}

\bibitem{Bethe31} H. Bethe, Z. Phys. {\bf 71}, 205 (1931).

\bibitem{Haldane83} F. D. M. Haldane, Phys. Lett. {\bf 93}, 464 (1983).

\bibitem{Dagotto95} E. Dagotto and T. M. Rice, Science {\bf 271}, 618 (1996).

\bibitem{Sachdev94} S. Sachdev, T. Senthil, and R. Shankar, Phys. Rev. B
{\bf 50}, 258 (1994).

\bibitem{Chitra} R. Chitra and T. Giamarchi, Phys. Rev. B {\bf 55}, 5816
(1997).

\bibitem{furusaki} A. Furusaki and S.--C. Zhang, Phys. Rev. B {\bf 60}, 1175
(1999).

\bibitem{Giamarchi99} T. Giamarchi and A. M. Tsvelik, Phys. Rev. B {\bf 59},
11398 (1999).

\bibitem{Giamarchi08} T. Giamarchi, Ch. R\"uegg, and O. Tchernyshyov, Nature
Physics {\bf 4}, 198 (2008), and references therein.

\bibitem{Schulz96} H. J. Schulz, Phys. Rev. Lett. {\bf 77}, 2790 (1996).

\bibitem{Lake05} B. Lake {\it et al.}, Nature Materials {\bf 4}, 329 (2005).

\bibitem{B_CuHpCl} M. Cl\'emancey {\it et al.}, Phys. Rev. Lett. {\bf 97},
167204 (2006).

\bibitem{Patyal} B. R. Patyal {\it et al.}, Phys. Rev. B
{\bf 41}, 1657 (1990).

\bibitem{Watson} B. C. Watson {\it et al.}, Phys. Rev. Lett. {\bf 86}, 5168
(2001).

\bibitem{Lorenz} T. Lorenz {\it et al.}, Phys. Rev. Lett. {\bf 100}, 067208
(2008).

\bibitem{Ruegg_specificheat} Ch. R\"uegg {\it et al.}, Phys. Rev. Lett. 
(in press).

\bibitem{Berthier_NMR} M. Klanjsek {\it et al.},  Phys. Rev. Lett. {\bf 101},
137207 (2008).

\bibitem{Thielemann} B. Thielemann {\it et al.}, (unpublished).

\bibitem{Nohadani} O. Nohadani, S. Wessel, B. Normand, and S. Haas, Phys.
Rev. B {\bf 69}, 220402(R) (2004).

\bibitem{nu_hs} The exponent $\nu$ cannot be determined quantitatively at
$B_s$ because of strong contributions to the heat capacity from the Schottky
anomaly of the nuclear spins \cite{Ruegg_specificheat}. The line in Fig.~1
is drawn with $\nu = 2$.

\bibitem{Oshikawa} Y. Maeda, C. Hotta, and M. Oshikawa, Phys. Rev. Lett.
{\bf 99}, 057205 (2007).

\bibitem{Matsumoto04} M. Matsumoto, B. Normand, T. M. Rice, and M. Sigrist,
Phys. Rev. B {\bf 69}, 054423 (2004).

\bibitem{Wessel} S. Wessel, M. Olshanii, and S. Haas, Phys. Rev. Lett.
{\bf 87}, 206407 (2001).

\bibitem{Yasuda05} C. Yasuda {\it et al.}, Phys. Rev. Lett. {\bf 94},
217201 (2005).

\bibitem{ALPS} A. F. Albuquerque {\it et al.}, J. Magn. Magn. Mat. {\bf 310},
1187 (2007).

\bibitem{ALPS_SSE} F. Alet {\it et al.}, Phys. Rev. E {\bf 71}, 036706,
(2005).

\end{thebibliography}
\end{document}